# Acoustic gravitational interaction revised

Ion Simaciu[1a], Zoltan Borsos[2] and Gheorghe Dumitrescu[3]

[1] Retired lecturer, Petroleum-Gas University of Ploiești, Ploiești 100680, Romania

[2] Petroleum-Gas University of Ploiești, Ploiești 100680, Romania

[3] Retired professor, High School Toma N. Socolescu, Ploiești, Romania

E-mail: [a] <isimaciu@yahoo.com>

**Abstract:** In this paper we deduce the expression of the gravito-acoustic force between two oscillating bubbles using the hypothesis that this type of force is a force of scattering-absorption of the energy of excitatory waves. The expression of the gravito-acoustic force at resonance highlights the dependence of this force on the product of the virtual masses of the two bubbles and on an acoustic gravitational constant. The acoustic gravitational constant depends on the absorption damping coefficient $\beta_{a0}$. We may say also that the expression of the acoustic gravitational constant is analogous to the expression of the gravitational constant in the electromagnetic world, that one obtained in the Einstein-Sciama model and the Dirac-Eddington large numbers hypothesis. The results obtained for this type of phenomenon in the acoustic world support the similarity of the acoustic world and the electromagnetic world.

Keyword: secondary Bjerknes force, gravito-acoustic force, the acoustic world

## 1. Introduction

In our previous paper [1, 2] we showed that the gravitational interaction can be modeled as a scatter-absorption of the CZPF background. In this paper, we return to the issue of the acoustic gravitational interaction that we tried to solve in [3]. We apply the same hypothesis, i.e., the gravito-acoustic interaction is a scattering-absorption interaction of the excitation waves. The new expression of gravito-acoustic force is better related to the expression of electro-acoustic force taking into account the analogy between the acoustic and electromagnetic worlds.

In the second part of this paper we infer the expression of the scattering-absorption force for two different bubbles and for identical bubbles. In the third section we work out the gravitational acoustic force when the bubbles are in a thermal background of acoustic waves. The fourth section presents the conclusions and discussions on this paper.

## 2. Gravito-acoustic force as a force of interaction by scattering-absorption

### 2.1. Deduction of the expression of the acoustic forces for scattering and absorption process

To infer the expression of the acoustic forces for scattering and absorption process, we start from the expression of the Bjerknes secondary force [4]

$$F_{B12}(r,\varphi) = -\frac{2\pi\rho\omega^2 R_{01}^3 R_{02}^3}{r^2} a_1 a_2 \cos(\varphi_2 - \varphi_1), \tag{1}$$

where:

$$a_i = \frac{A}{\rho R_{0i}^2 \sqrt{(\omega^2 - \omega_{0i}^2)^2 + 4\beta_i^2 \omega^2}}, \quad \varphi_i = \arctan\frac{2\beta_i \omega}{(\omega^2 - \omega_{0i}^2)}, i = 1, 2 \tag{2}$$

and so:

$$\cos\varphi_i = \frac{(\omega^2 - \omega_i^2)}{\sqrt{(\omega^2 - \omega_i^2)^2 + 4\beta_i^2\omega^2}}, \quad \sin\varphi_i = \frac{2\beta_i\omega}{\sqrt{(\omega^2 - \omega_i^2)^2 + 4\beta_i^2\omega^2}}, i = 1,2;$$

$$\cos(\varphi_2 - \varphi_1) = \cos\varphi_2 \cos\varphi_1 + \sin\varphi_2 \sin\varphi_1 = \frac{(\omega^2 - \omega_1^2)(\omega^2 - \omega_2^2) + 4\beta_1\beta_2\omega^2}{\sqrt{(\omega^2 - \omega_1^2)^2 + 4\beta_1^2\omega^2}\sqrt{(\omega^2 - \omega_2^2)^2 + 4\beta_2^2\omega^2}}. \tag{3}$$

Replacing $\cos(\varphi_2 - \varphi_1)$ from Eqs. (3) and the amplitudes of the oscillations from Eq. (2) in the expression of the force (1), results

$$F_{B12}(r) = -\frac{2\pi R_1 R_2 \omega^2 A^2 \left[(\omega^2 - \omega_1^2)(\omega^2 - \omega_2^2) + 4\beta_1\beta_2\omega^2\right]}{r^2 \rho \left[(\omega^2 - \omega_1^2)^2 + 4\beta_1^2\omega^2\right]\left[(\omega^2 - \omega_2^2)^2 + 4\beta_2^2\omega^2\right]}. \tag{4}$$

The damping coefficient has three components [4]:

$$\beta_i = \beta_{aci} + \beta_{vi} + \beta_{thi} = \beta_{aci} + (\beta_{vi} + \beta_{thi}) = \beta_{si} + \beta_{ai} = \beta_{si}\left(1 + \frac{\beta_{ai}}{\beta_{si}}\right), \beta_{ai} \ll \beta_{si}, \tag{5}$$

with:

$$\beta_{aci} = \beta_{si} = \frac{\omega^2 R_i}{2u}, \quad \beta_{vi} + \beta_{thi} = \frac{2\mu}{\rho R_i^2} + \frac{2\mu_{thi}}{\rho R_i^2} = \beta_{ai}, \tag{6}$$

that is, the damping of the oscillations is induced by the absorption of the oscillation energy in the liquid as an effect of viscosity (the coefficient $\beta_{vi}$) and by the absorption of the oscillation energy in the gas or/and vapor inside the bubbles as an effect of compression and decompression (the coefficient $\beta_{thi}$) of the gas or/and vapor.

Replacing Eq. (6) in Eq. (4) yields,

$$F_{B12}(r) = -\frac{2\pi R_1 R_2 \omega^2 A^2 \left[(\omega^2 - \omega_1^2)(\omega^2 - \omega_2^2) + 4(\beta_{s1} + \beta_{a1})(\beta_{s2} + \beta_{a2})\omega^2\right]}{r^2 \rho \left[(\omega^2 - \omega_1^2)^2 + 4(\beta_{s1} + \beta_{a1})^2 \omega^2\right]\left[(\omega^2 - \omega_2^2)^2 + 4(\beta_{s2} + \beta_{a2})^2 \omega\right]} \cong$$

$$\frac{-2\pi R_1 R_2 \omega^2 A^2 \left[(\omega^2 - \omega_1^2)(\omega^2 - \omega_2^2) + 4\omega^2 \beta_{s1}\beta_{s2}\right]}{r^2 \rho \left[(\omega^2 - \omega_1^2)^2 + 4\beta_{s1}^2\omega^2\right]\left[(\omega^2 - \omega_2^2)^2 + 4\beta_{s2}^2\omega^2\right]} +$$

$$\frac{-8\pi R_1 R_2 \omega^4 A^2 (\beta_{s1}\beta_{a2} + \beta_{s2}\beta_{a1})}{r^2 \rho \left[(\omega^2 - \omega_1^2)^2 + 4\beta_{s1}^2\omega^2\right]\left[(\omega^2 - \omega_2^2)^2 + 4\beta_{s2}^2\omega^2\right]} + \tag{7}$$

$$\frac{-8\pi R_1 R_2 \omega^4 A^2 \beta_{a1}\beta_{a2}}{r^2 \rho \left[(\omega^2 - \omega_1^2)^2 + 4\beta_{s1}^2\omega^2\right]\left[(\omega^2 - \omega_2^2)^2 + 4\beta_{s2}^2\omega^2\right]} = F_{B12ss}(r) + F_{B12sa}(r) + F_{B12aa}(r).$$

We kept in Eq. (7) only the terms, $\beta_{si}\beta_{aj}, i \neq j$, of the first order.

$F_{B12ss}(r)$ is the acoustic force component for the scattering-scattering process, which we will call the electro-acoustic force

$$F_{B12ss}(r) = \frac{-2\pi R_1 R_2 \omega^2 A^2 \left[(\omega^2 - \omega_1^2)(\omega^2 - \omega_2^2) + 4\omega^2 \beta_{s1}\beta_{s2}\right]}{r^2 \rho \left[(\omega^2 - \omega_1^2)^2 + 4\beta_{s1}^2\omega^2\right]\left[(\omega^2 - \omega_2^2)^2 + 4\beta_{s2}^2\omega^2\right]}, \qquad (8)$$

$F_{B12sa}(r)$ is the acoustic force component for the scattering-absorbtion process, which we will call the gravito-acoustic force

$$F_{B12sa}(r) = \frac{-8\pi R_1 R_2 \omega^4 A^2 (\beta_{s1}\beta_{a2} + \beta_{s2}\beta_{a1})}{r^2 \rho \left[(\omega^2 - \omega_1^2)^2 + 4\beta_{s1}^2\omega^2\right]\left[(\omega^2 - \omega_2^2)^2 + 4\beta_{s2}^2\omega^2\right]} \qquad (9)$$

and $F_{B12saa}(r)$ is the acoustic force component for the absorption-absorption process, that is much smaller than the gravito-acoustic one

$$F_{B12aa}(r) = \frac{-8\pi R_1 R_2 \omega^4 A^2 \beta_{a1}\beta_{a2}}{r^2 \rho \left[(\omega^2 - \omega_1^2)^2 + 4\beta_{s1}^2\omega^2\right]\left[(\omega^2 - \omega_2^2)^2 + 4\beta_{s2}^2\omega^2\right]}. \qquad (10)$$

For identical bubbles, $\omega_1 = \omega_2 = \omega_0 = (p_{ef}/\rho)^{1/2} R_0^{-1}$ and $\beta_s = \omega^2 R_0/(2u)$, the expressions of the three forces mentioned above become:

$$F_{Bss}(r) = \frac{-2\pi R_0^2 \omega^2 A^2}{r^2 \rho \left[(\omega^2 - \omega_0^2)^2 + 4\beta_s^2\omega^2\right]} = \frac{-2\pi R_0^2 \omega^2 A^2}{r^2 \rho \left[(\omega^2 - \omega_0^2)^2 + R_0^2\omega^6/u^2\right]}, \qquad (11)$$

$$F_{Bsa}(r) = \frac{-16\pi R_0^2 \omega^4 A^2 \beta_s \beta_a}{r^2 \rho \left[(\omega^2 - \omega_0^2)^2 + 4\beta_s^2\omega^2\right]^2} = \frac{-8\pi R_0^3 \omega^6 A^2 \beta_a}{r^2 \rho u \left[(\omega^2 - \omega_0^2)^2 + R_0^2\omega^6/u^2\right]^2}, \qquad (12)$$

$$F_{Baa}(r) = \frac{-8\pi R_0^2 \omega^4 A^2 \beta_a^2}{r^2 \rho \left[(\omega^2 - \omega_0^2)^2 + R_0^2\omega^6/u^2\right]^2}. \qquad (13)$$

The expression of the electro-acoustic force (11) is the same as that one obtained in our previous papers (Eq. (20) of paper [5] or Eq: (2) of paper [6]).

## 2.2. Properties of gravito-acoustic force, gravito-acoustic constant

The gravito-acustic force is much smaller than the corresponding electro-acoustic force (Eq. (20) of [5] or Eq: (2) of [6]). The ratio of these forces is

$$\frac{F_{Bsa}(r)}{F_{Bss}(r)} = \frac{\dfrac{-8\pi R_0^3 \omega^6 A^2 \beta_a}{r^2 \rho u \left[(\omega^2 - \omega_0^2)^2 + R_0^2\omega^6/u^2\right]^2}}{\dfrac{-2\pi R_0^2 \omega^2 A^2}{r^2 \rho \left[(\omega^2 - \omega_0^2)^2 + R_0^2\omega^6/u^2\right]}} = \frac{4 R_0 \omega^4 \beta_a}{\left[(\omega^2 - \omega_0^2)^2 + R_0^2\omega^6/u^2\right] u}. \qquad (14)$$

At resonance [4], $\omega \to \omega_0 = (p_{ef}/\rho)^{1/2} R_0^{-1}$, the gravito-acoustic force (12) becomes

$$F_{B0sa}(r) = \frac{-16\pi R_0^3 \omega_0^6 A^2 \beta_{a0}}{r^2 \rho (R_0^2 \omega_0^6/u^2)^2} = \frac{-8\pi A^2 \beta_{a0} u^3}{r^2 \rho \omega_0^6 R_0} = \frac{-8\pi R_0^5 \rho^2 u^3 A^2 \beta_{a0}}{r^2 p_{ef}^3} \qquad (15)$$

and the ratio mentioned above becomes

$$\lim_{\omega \to \omega_0}\left(\frac{F_{B0sa}(r)}{F_{B0ss}(r)}\right) = \lim_{\omega \to \omega_0}\left\{\frac{4R_0\omega^4\beta_a}{\left[(\omega^2 - \omega_0^2)^2 + R_0^2\omega^6/u^2\right]u}\right\} = \frac{2\beta_{a0}}{\beta_{s0}} = \left(\frac{R_0}{u/\beta_{a0}}\right)\left(\frac{4\rho u^2}{p_{ef}}\right). \tag{16}$$

Since the absorption damping coefficient, $\beta_{a0}$, has the dimensions of a frequency, then the ratio $u/\beta_{a0}$ has the dimensions of a length. The parameter

$$\left(\frac{u}{\beta_{a0}}\right)\left(\frac{p_{ef}}{4\rho u^2}\right) = R_{a0} \tag{17}$$

is the acoustic length of the resonant absorption damping. With these, the ratio of the acoustic forces has an expression similar to the ratio of the gravitational force and the electrostatic force between two identical charged particles that interact, $F_g(r)/F_e(r) = Gm^2/e^2 = (Gm/c^2)/(e^2/mc^2) \cong R_g/R_e \cong R_e/R_u$, according to the Edington hypothesis [7],

$$\lim_{\omega \to \omega_0}\left(\frac{F_{B0sa}(r)}{F_{B0ss}(r)}\right) = \frac{R_0}{R_{a0}} = \frac{R_{as0}}{R_0}. \tag{18}$$

With the virtual mass of the bubble $m_b = 2\pi R_0^3 \rho/3$ [4], the expression (15) of gravito-acoustic force, at resonance, becomes

$$F_{B0sa}(r) = \frac{-8\pi R_0^5 A^2 \rho^2 \beta_{a0} u^3}{r^2 p_{ef}^3} = \left(\frac{-4\pi^2 R_0^6 \rho^2}{9\pi r^2}\right)\left[\left(\frac{A^2 \rho u^2}{p_{ef}^3}\right)\left(\frac{18\pi \beta_{a0} u}{\rho R_0}\right)\right] = \left(\frac{-m_b^2}{r^2}\right)\left[\left(\frac{18 A^2 \rho u^2}{\pi p_{ef}^3}\right)\left(\frac{\beta_{a0} u}{\rho R_0}\right)\right]. \tag{19}$$

According to Eq. (19), the expression of the gravito-acoustic force, between two identical bubbles, at resonance, is proportional to the product of the virtual masses of the two bubbles and inversely proportional to the distance, i.e., it is similar to the Newtonian expression of the gravitational force $F_N(r) = Gm^2/r^2$ [8, Sch. 7.6].

The constant of the gravito-acoustic interaction (the gravito-acoustic constant) depends on the absorption damping coefficient, $\beta_{a0} = \beta_a(\omega_0)$, and also on the bubble and the liquid parameters

$$G_{0a} = \left(\frac{18 A^2 \rho u^2}{\pi p_{ef}^3}\right)\left(\frac{\beta_{a0} u}{\rho R_0}\right). \tag{20a}$$

Replacing the absorption damping coefficient, $\beta_{a0} = \beta_a(\omega_0)$, (17), and the virtual mass, $m_b = 2\pi R_0^3 \rho/3$, in the expression of the gravito-acoustic constant (20a), yields

$$G_{0a} = \left(\frac{3A^2}{p_{ef}^2}\right)\left(\frac{R_0^2 u^2}{R_{a0} m_b}\right). \tag{20b}$$

Interestingly, this expression of the gravito-acoustic constant is compatible with the expression of the gravitational constant in the electromagnetic world obtained in the Einstein-Sciama model [9] and the Dirac-Eddington large numbers hypothesis [7, 10].

$$G \cong \frac{c^2 R_e^2}{m R_U}. \tag{21}$$

The resemblance of the two relations further strengthens the analogy between the two "worlds." As we can see, (21) relates the radius of the finite universe, $R_U$, to the gravitational radius, $R_g$, of the constituent elementary particles of the universe (electron, proton or a charged generic particle with mass equal to their average mass) , $R_U \cong N R_g$, [9] and also is similar to the Dirac-Eddington relation which relates the radius of the universe, $R_U \cong \sqrt{N} R_e$, to the electromagnetic radius, $R_e$, of the same fundamental particles [7].

Merging the above relations yields: $R_e \cong \sqrt{N} R_g$ or $R_g \cong R_e^2 / R_U$ or $Gm/c^2 \cong R_e^2 / R_U$ and therefore, implicitly the expression of the gravitational constant, Eq. (21), in the electromagnetic universe/ world. It depends on the radius of the universe, the electromagnetic radius, and the mass of the particle.

The similarity mentioned above is possible when in the acoustic universe one emphasize that, when several bubbles that form a cluster, each bubble being excited by a background of acoustic waves, the background has such a high average pressure $A_f$ so that the constant $3 A_f^2 / p_{ef}^2 \cong 1$ is close to unity.

Also, by analyzing the acoustic phenomena in a cluster it will lead to the significance of the acoustic length of damping process, by absorption at resonance, and its connection with the radius of the finite universe. We will investigate these hypotheses and connections in the acoustic phenomena of a cluster in further works.

Substituting the damping coefficient $\beta_{a0}$ given by (6) in (20a), yields

$$G_{0a} = \left( \frac{A^2 \rho u^2}{p_{ef}^3} \right) \left( \frac{18 \beta_{a0} u}{\pi \rho R_0} \right) = \left( \frac{A^2 \rho u^2}{p_{ef}^3} \right) \left[ \frac{18u}{\pi \rho R_0} (\beta_{v0} + \beta_{th0}) \right] = \\ \left( \frac{A^2 \rho u^2}{p_{ef}^3} \right) \left[ \frac{18u}{\pi \rho R_0} \left( \frac{2\mu}{\rho R_0^2} \right) \right] + \left( \frac{A^2 \rho u^2}{p_{ef}^3} \right) \left( \frac{18 \beta_{th0} u}{\pi \rho R_0} \right) = G_{0av} + G_{0ath} \qquad (22)$$

that is, the gravito-acoustic interaction is produced both by the absorption of acoustic energy in the liquid and in the gas and/or vapor of the inner bubble. If we replace the coefficient (6) in the expression of the acoustic resonance damping length, Eq. (17), and defining the liquid damping length and the thermal damping length as

$$R_{v0} = \left( \frac{u}{\beta_{v0}} \right) \left( \frac{p_{ef}}{4 \rho u^2} \right), R_{v0} = \left( \frac{u}{\beta_{th0}} \right) \left( \frac{p_{ef}}{4 \rho u^2} \right), \qquad (23)$$

one obtains the relationship between the three lengths of acoustic absorption damping

$$R_{a0} = \frac{R_{v0} R_{th0}}{R_{v0} + R_{th0}}. \qquad (24)$$

In a superfluid [11], the viscosity coefficient is zero, $\mu = 0$, $R_{v0} \to \infty$, $R_{a0} \to R_{th0}$, and therefore the gravito-acoustic interaction is a function only of the absorption of energy in the gas or vapor of the inner bubbles

$$G_{a0sf} = G_{a0th} = \left( \frac{9 A^2}{2 \pi p_{ef}^2} \right) \left( \frac{u^2}{\rho R_0 R_{th0}} \right) = \left( \frac{3 A^2}{p_{ef}^2} \right) \left( \frac{u^2 R_0^2}{m_b R_{th0}} \right). \qquad (25)$$

We will address the modeling of acoustic phenomena in superfluid in future papers.

## 3. Gravito-acoustic force in the presence of a background of acoustic waves

In what it follows we will express the gravitational acoustic force when the bubbles are in a thermal background of acoustic waves. To do it, we will use the method applied in [5, 6] and also we will use the expression of the acoustic gravitational force in a background of acoustic waves. For the thermal acoustic background, the spectral densities, $\rho(\omega,T) = dU/(Vd\omega)$, Eq. (25) of Chapter 5 of [12], has the form

$$\rho(\omega,T) = \frac{\hbar\omega^3}{2\pi^2 u^3}\frac{1}{\exp[\hbar\omega/(kT)]-1}. \tag{26}$$

The expression of the gravito-acoustic force, between two identical bubbles, in a thermal background of waves and for the fraction of waves with angular frequency $\omega$, is

$$\delta F_{Bsa}(r,T) = \frac{-8\pi R_0^3 \omega^6 (\delta A)(\delta A')\beta_a}{r^2 \rho u \left[(\omega^2-\omega_0^2)^2 + R_0^2\omega^6/u^2\right]^2}, \tag{27}$$

which is similar to Eq. (43) of paper [6] or Eq. (64) of paper [5].

With the elementary amplitude given by the expression (25) in the paper [6], $\delta A = (p_0\gamma_f \omega/u)\cdot h(\omega,\theta,T)\exp\left[-i(\omega t - \vec{k}\vec{r} - \theta)\right]d^3k$, Eq. (27) becomes

$$\delta F_{Bsa}(r,T) = \frac{-8\pi R_0^3 \omega^8 \beta_a (p_0\gamma_f)^2}{r^2 \rho u^3}\left\{\frac{h(\omega,\theta,T)\exp\left[-i(\omega t - \vec{k}\vec{r} - \theta - \varphi)\right]d^3k}{\left[(\omega^2-\omega_0^2)^2 + R_0^2\omega^6/u^2\right]}\right\}\times$$

$$\left\{\frac{h(\omega',\theta',T)\exp\left[-i(\omega' t - \vec{k}'\vec{r} - \theta - \varphi)\right]d^3k'}{\left[(\omega'^2-\omega_0^2)^2 + R_0^2\omega'^6/u^2\right]}\right\}. \tag{28}$$

One can see in Eq. (28) that the amplitude of the oscillation of the bubbles contains the phase oscillation $\varphi$ also to agree to Eq. (38) from [6]. By integrating and performing the average for random phase in Eq. (28), the gravito-acoustic force has the form

$$F_{Bsa}(r,T) = \frac{-8\pi R_0^3 (p_0\gamma_f)^2}{r^2 \rho u^3}\text{Re}\left\langle\iint\frac{\left\{\omega^8\beta_a h(\omega,\theta,T)\exp\left[-i(\omega t - \vec{k}\vec{r} - \theta - \varphi)\right]d^3k\right\}}{\left[(\omega^2-\omega_0^2)^2 + R_0^2\omega^6/u^2\right]}\times\right.$$

$$\left.\left\{\frac{h(\omega',\theta',T)\exp\left[i(\omega' t - \vec{k}'\vec{r} - \theta - \varphi)\right]d^3k'}{\left[(\omega'^2-\omega_0^2)^2 + R_0^2\omega'^6/u^2\right]}\right\}\right\rangle = \frac{-8\pi R_0^3 (p_0\gamma_f)^2}{r^2 \rho u^3}\times \tag{29}$$

$$\int_0^\infty \frac{\omega^8\beta_a h^2(\omega,\theta,T)d^3k}{\left[(\omega^2-\omega_0^2)^2 + R_0^2\omega^6/u^2\right]^2} = \frac{-32\pi^2 R_0^3 (p_0\gamma_f)^2}{r^2 \rho u^6}\int_0^\infty\frac{\beta_a h^2(\omega,\theta,T)\omega^{10}d\omega}{\left[(\omega^2-\omega_0^2)^2 + R_0^2\omega^6/u^2\right]^2}.$$

In Eq. (29), the parenthesis, $\langle\ \rangle$, symbolize an average after the random phase.

According to Eq. (37) of the paper [6], the amplitude of the excitation wave is

$$h(\omega,T) = \frac{\hbar^{1/2}}{\left[4\pi^3 \rho \omega \left(\exp\frac{\hbar\omega}{kT} - 1\right)\right]^{1/2}}. \tag{30}$$

Substituting Eq. (30) into Eq. (29), we finally find

$$F_{Bsa}(r,T) = \frac{-8R_0^3 \hbar (p_0 \gamma_f)^2}{\pi r^2 \rho^2 u^6} \int_0^\infty \frac{\beta_a \omega^9 d\omega}{\left(\exp\frac{\hbar\omega}{kT} - 1\right)\left[(\omega^2 - \omega_0^2)^2 + R_0^2 \omega^6/u^2\right]^2} = \frac{-8R_0^3 \hbar (p_0 \gamma_f)^2}{\pi r^2 \rho^2 u^6} I_\omega. \tag{31}$$

To estimate analytically the integral of Eq. (31) we use the saddle-point method [8-Ch. 23, 13]. First we make the change of variable $\omega - \omega_0 = y$. With this change of variable yields: $d\omega = dy$, $\omega + \omega_0 \cong 2\omega_0$, $\beta_a(\omega) = \beta_a(\omega_0) = \beta_{a0}$, $\exp[\hbar\omega/(kT)] - 1 = \exp[\hbar\omega_0/(kT)] - 1$ and $\omega^n \to \omega_0^n$ Substituting in the expression of the integral it becomes

$$I_\omega = \int_0^\infty \frac{\beta_a \omega^9 d\omega}{\left(\exp\frac{\hbar\omega}{kT} - 1\right)\left[(\omega^2 - \omega_0^2)^2 + R_0^2 \omega^6/u^2\right]^2} = I_y \cong$$

$$\frac{\beta_{a0} \omega_0^9}{\left(\exp\frac{\hbar\omega_0}{kT} - 1\right)} \int_{-\omega_0}^\infty \frac{dy}{16\omega_0^4 \left[y^2 + (R_0 \omega_0^2/2u)^2\right]^2} = \tag{32}$$

$$\frac{\beta_{a0} \omega_0^5}{16\left(\exp\frac{\hbar\omega_0}{kT} - 1\right)} \int_{-\omega_0}^\infty \frac{dy}{\left[y^2 + (R_0 \omega_0^2/2u)^2\right]^2} = \frac{\pi \beta_{a0} u^3}{8\left(\exp\frac{\hbar\omega_0}{kT} - 1\right) R_0^3 \omega_0}.$$

Replacing the integral (32) in the expression force (5) one obtains

$$F_{Bsa}(r,T) = \frac{-\hbar u}{r^2 \left(\exp\frac{\hbar\omega_0}{kT} - 1\right)} \left(\frac{p_0 \gamma_f}{\rho u^2}\right)^2 \frac{\beta_{a0}}{\omega_0}. \tag{33}$$

Taking into account the expression of the electro-acoustic force in a thermal background, Eq. (71), from paper [5] or Eq. (49), from paper [6], then results an expression of the gravito-acoustic force to be a function of the electro-acoustic force between the same bubbles:

$$F_{Bsa}(r,T) = \left[\frac{-\hbar\omega_0 R_0}{r^2 \left(\exp\frac{\hbar\omega_0}{kT} - 1\right)} \left(\frac{p_0 \gamma_f}{\rho u^2}\right)^2\right] \frac{u\beta_{a0}}{R_0 \omega_0^2} = \frac{u\beta_{a0}}{R_0 \omega_0^2} F_{Bss}(r,T) = \frac{\beta_{a0}}{2\beta_{s0}} F_{Bss}(r,T) \tag{34}$$

or

$$\frac{F_{Bsa}(r,T)}{F_{Bss}(r,T)} = \frac{\beta_{a0}}{2\beta_{s0}}. \tag{35}$$

This ratio of gravito-acoustic and electro-acoustic forces is similar to the ratio of the absorption force (gravitational force) and the scattering force (electrostatic force) in the electromagnetic world/universe. That is, the above ratio between the gravitational force and the electrostatic force between two identical particles tacking place for: $m_1 = m_2 = m$, $\Gamma'_1 = \Gamma'_2 = \Gamma'$, $R_1 = R_2 = R$, $\omega_{12} = \omega_0$ and $\Gamma = 2\beta$, according to Eq. (10) of papers [1, 2],

$$F_{sa}(r) = \frac{3\hbar c}{r^2}\frac{\Gamma'(\omega_0)}{\omega_0} = \frac{\hbar \omega_0 R_0}{r^2}\frac{3c\Gamma'(\omega_0)}{\omega_0^2 R_0} \cong F_{ss}(r)\frac{\Gamma'(\omega_0)}{\Gamma(\omega_0)} \text{ or } \frac{F_{sa}(r)}{F_{ss}(r)} = \frac{\Gamma'(\omega_0)}{\Gamma(\omega_0)} = \frac{\beta'(\omega_0)}{\beta(\omega_0)}. \quad (36)$$

We find the same ratio, $F_N(r)/F_C(r) = Gm^2/e^2 = Gmc^2/(e^2c^2/m) \cong R_g/R_e$ if we make the replacement $R_g/R_e = \omega_0^2 R_g c/(\omega_0^2 R_e c) = \Gamma'/\Gamma = \beta'(\omega_0)/\beta(\omega_0)$ according to the hypothesis that $\Gamma' \sim R_g \omega_0^2/c$ (Eq. (15) of the papers [1, 2].

## 4. Conclusions and discussions

Assuming the hypothesis that the gravito-acoustic force is a force of scattering-absorption of the excitatory wave, we inferred the expression of this type of acoustic force between two bubbles. This force is proportional to the absorption damping coefficient $\beta_a$.

For identical bubbles, at resonance, the expression of the gravito-acoustic force facilitated to express the constant of gravito-acoustic interaction $G_{0a}$. This constant depends both on the absorption of the oscillation energy of the bubble in the liquid, by means of the coefficient $\beta_{v0}$ and on the absorption of the energy in the gas or/and vapor of inner bubbles, by means of the coefficient $\beta_{th0}$. In a superfluid, the gravito-acoustic force is a function of only the absorption of energy in the gas or/and vapor of inner bubbles. We consider that this relation is important on the horizon that the vacuum can be modeled as a physical environment with superfluid and superconducting properties [14, 15].

The expression of the gravito-acoustic constant is, according to Eq. (20), analogous to the expression of the gravitational constant for a finite universe model Einstein-Sciama and the Dirac-Edington large numbers hypothesis, if we introduce the magnitude of the acoustic length of damping by absorption at resonance, $R_{a0} \sim u/\beta_{a0}$.

The significance of this quantity will be clearer when we will study the form of the gravito-acoustic force between two bubbles inside a cluster. The form of the expression of the gravito-acoustic force in a thermal acoustic background will allow maybe to find out its ratio to the electro-acoustic force. This ratio is, up to a small numerical constant, identical to that of the ratio of the two acoustic forces to resonance and analogous to that between the gravitational force and the electrostatic force. The results obtained in this paper confirm the analogy between the acoustic world and the electromagnetic world.